\documentclass{svjour3}

\usepackage{cite}

\usepackage{multirow}
\usepackage{dcolumn}
\usepackage{graphicx}
\usepackage{color}

\begin{document}

\title{Eigenvalues and eigenfunctions of the anharmonic oscillator $%
V(x,y)=x^{2}y^{2}$}

\author{Francisco M. Fern\'{a}ndez \and Javier Garcia}

\institute{Francisco M Fern\'andez \and Javier Garcia \at INIFTA
(UNLP, CCT La Plata--CONICET), Divisi\'{o}n Qu\'{i}mica
Te\'{o}rica Diag. 113 y 64 (S/N), Sucursal 4, Casilla de Correo 16
1900 La Plata, Argentina \email{fernande@quimica.unlp.edu.ar}}

\date{Received: date / Accepted: date}

\maketitle

\begin{abstract}
We obtain sufficiently accurate eigenvalues and eigenfunctions for the
anharmonic oscillator with potential $V(x,y)=x^{2}y^{2}$ by means of three
different methods. Our results strongly suggest that the spectrum of this
oscillator is discrete in agreement with early rigorous mathematical proofs
and against a recent statement that cast doubts about it.
\end{abstract}



\section{Introduction}

Some time ago Bender et al\cite{BDMS01} stated that it is not known if the
spectrum of the anharmonic oscillator potential $V(x,y)=x^{2}y^{2}$ is
discrete. Several years earlier Simon\cite{S83} had given five proofs that
the spectrum of such oscillator is indeed discrete.

We are not aware of any calculation of the eigenvalues and
eigenfunctions of that anharmonic oscillator. For this reason we
will provide some reasonably accurate results in this paper. In
section~\ref{sec:RRVM} we outline the application of the
Rayleigh-Ritz variational method taking into account the
point-group symmetry of the oscillator. In
section~\ref{sec:Moments} we discuss two approaches based on the
moments of the Hamiltonian operator: the Rayleigh-Ritz method in
the Krylov space (RRK)\cite{AF09} (and references therein) and the
connected-moments expansion (CMX)\cite{C87e,K87}. In
section~\ref{sec:Results} we compare and discuss the results
obtained by the three approaches and draw conclusions.

\section{Rayleigh-Ritz variational method}

\label{sec:RRVM}

As stated in the introduction, we are interested in the eigenvalues and
eigenfunctions of the anharmonic oscillator
\begin{equation}
H=p_{x}^{2}+p_{y}^{2}+x^{2}y^{2}.  \label{eq:H}
\end{equation}
In this section we outline the application of the well known Rayleigh-Ritz
variational method. We choose products $\varphi _{mn}(x,y)=\phi _{m}(x)\phi
_{n}(y)$ of eigenfunctions $\phi _{n}(q)$, $n=0,1,\ldots $, of the harmonic
oscillator $H=p_{q}^{2}+q^{2}$ as a suitable basis set.

Like the Pullen-Edmonds Hamiltonian\cite{PE81a} the Hamiltonian (\ref{eq:H})
is invariant under the symmetry operations of the point group $C_{4v}$\cite
{C90,T64}. Therefore, the appropriate basis functions are
\begin{equation}
\begin{array}{ll}
\varphi _{2m\,2n}^{+}(x,y),\;m,n=0,1,\ldots & A_{1} \\
\varphi _{2m+1\,2n+1}^{-}(x,y),\;m\neq n=0,1,\ldots & A_{2} \\
\varphi _{2m\,2n}^{-}(x,y),\;m\neq n=0,1,\ldots & B_{1} \\
\varphi _{2m+1\,2n+1}^{+}(x,y),\;m,n=0,1,\ldots & B_{2} \\
\left\{ \varphi _{2m\,2n+1}(x,y),\varphi _{2m+1\,2n}(x,y)\right\}
,\;m,n=0,1,\ldots & E
\end{array}
,  \label{eq:HOPG}
\end{equation}
where
\begin{eqnarray}
\varphi _{mn}^{+}(x,y) &=&\frac{1}{\sqrt{2(1+\delta _{mn})}}\left( \varphi
_{mn}+\varphi _{nm}\right) ,  \nonumber \\
\varphi _{mn}^{-}(x,y) &=&\frac{1}{\sqrt{2}}\left( \varphi _{mn}-\varphi
_{nm}\right) .  \label{eq:basis_funct}
\end{eqnarray}
An obvious advantage of using point-group symmetry is that we diagonalize
the Hamiltonian matrix $\mathbf{H}^{S}$ for each irreducible representation $%
S=A_{1},A_{2},B_{1},B_{2},E$ separately. What is more: we can even split the
calculation for the two-dimensional irreducible representation $E$ into its
two components, which decreases the dimension of the matrices still further.
Thus, point-group symmetry simplifies all the calculations and enables us to
interpret the results more clearly. We will refer to this Rayleigh-Ritz
method with the harmonic-oscillator basis set as RRHO.

\section{Moments methods}

\label{sec:Moments}

In this section we discuss two methods based on the moments of the
Hamiltonian operator
\begin{equation}
\mu _{j}=\frac{\left\langle \varphi \right| H^{j}\left| \varphi
\right\rangle }{\left\langle \varphi \right| \left. \varphi \right\rangle },
\label{eq:moments}
\end{equation}
where $\varphi $ is a properly chosen reference function.

The first one is the Rayleigh-Ritz variational method in the Krylov space
(RRK) spanned by the non-orthogonal basis set of functions
\begin{equation}
f_{j}=H^{j}\varphi ,\;j=0,1,\ldots ,  \label{q:basis_K}
\end{equation}
which has been successfully applied to the Pullen-Edmonds Hamiltonian\cite
{AF09}.

The second approach is the connected-moments expansion (CMX) developed by
Cioslowski\cite{C87e} who tested it on the ground state of the
Pullen-Edmonds Hamiltonian. Amore and Fern\'{a}ndez\cite{AF09} carried out a
calculation of much larger order on the ground and excited states by means
of the compact and most elegant formula developed by Knowles\cite{K87} that
we also use in this paper.

For the application of both moments methods we resort to the following
reference functions
\begin{eqnarray}
\varphi _{A_{1}} &=&\exp \left( -a\left[ x^{2}+y^{2}\right] \right)
\nonumber \\
\varphi _{A_{2}} &=&xy(x^{2}-y^{2})\exp \left( -a\left[ x^{2}+y^{2}\right]
\right)  \nonumber \\
\varphi _{B_{1}} &=&(x^{2}-y^{2})\exp \left( -a\left[ x^{2}+y^{2}\right]
\right)  \nonumber \\
\varphi _{B_{2}} &=&xy\exp \left( -a\left[ x^{2}+y^{2}\right] \right)
\nonumber \\
\varphi _{E} &=&\left\{
\begin{array}{c}
x\exp \left( -a\left[ x^{2}+y^{2}\right] \right) \\
y\exp \left( -a\left[ x^{2}+y^{2}\right] \right)
\end{array}
\right. .
\end{eqnarray}
Note that the reference function $\varphi _{A_{2}}$ was not
considered in the application of these methods to the
Pullen-Edmonds Hamiltonian\cite{AF09}.

\section{Results and discussion}

\label{sec:Results}

\begin{figure}[H]
~\bigskip\bigskip
\par
\begin{center}
\includegraphics[width=6cm]{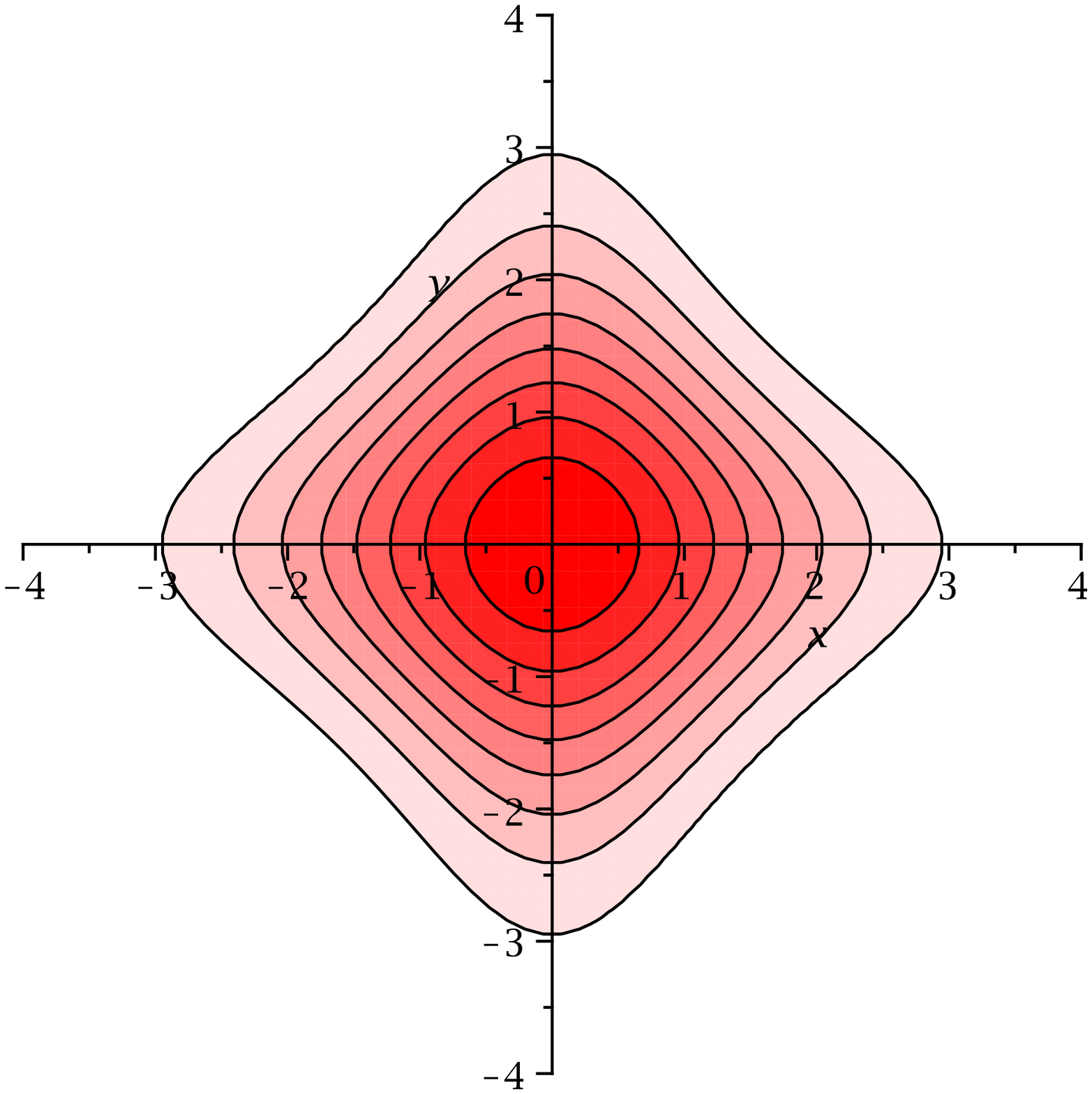} \includegraphics[width=6cm]{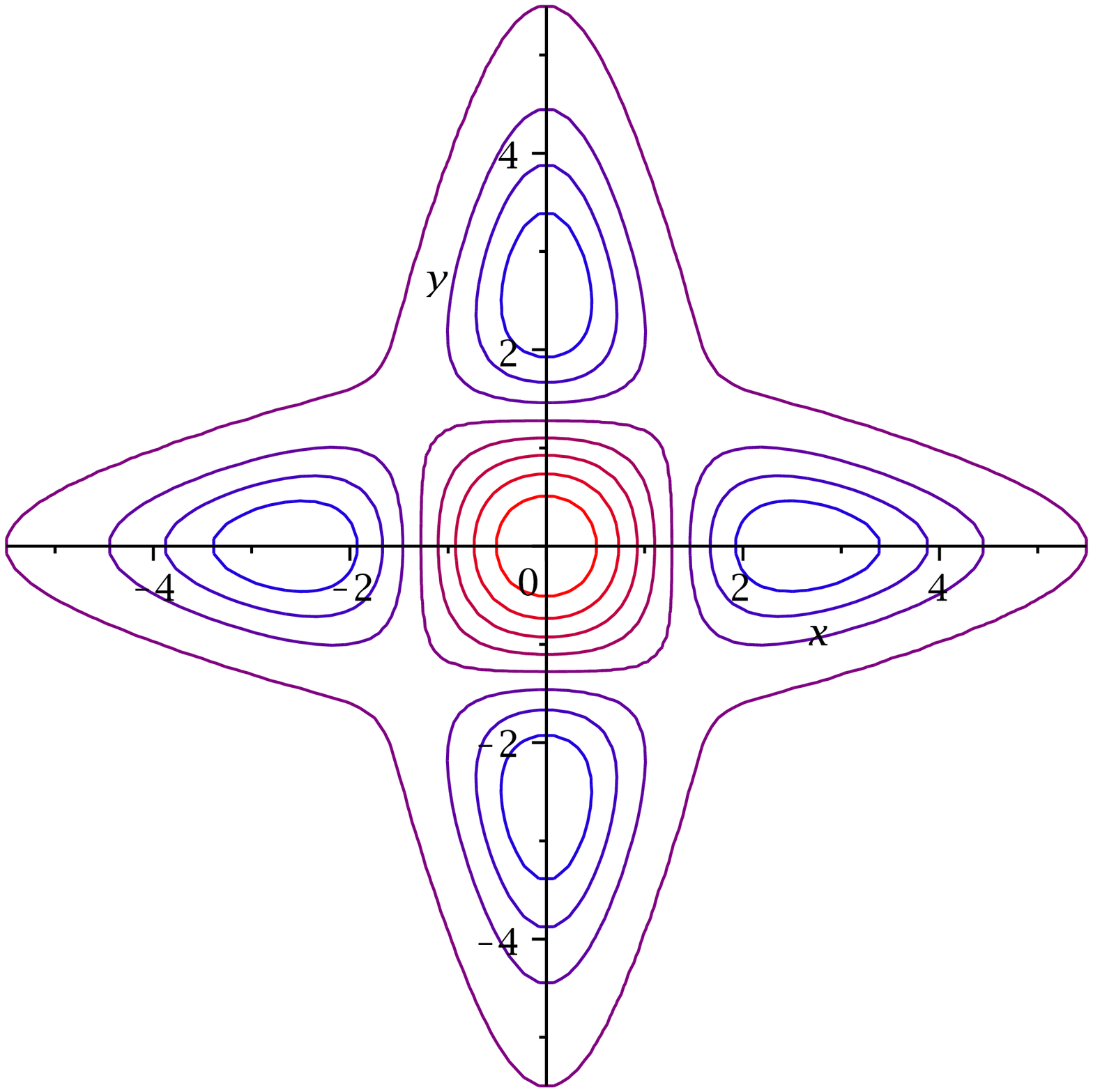} %
\includegraphics[width=6cm]{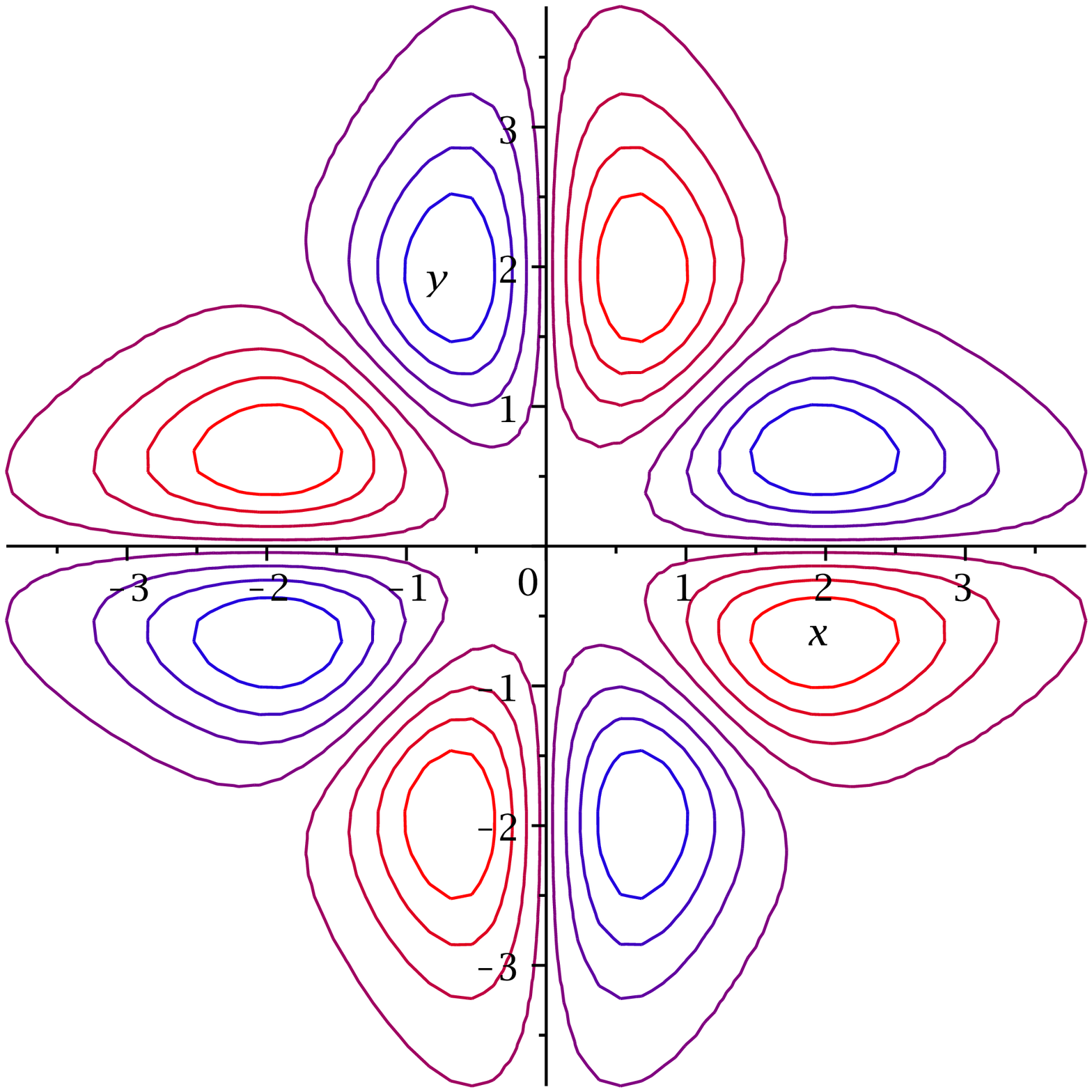} \includegraphics[width=6cm]{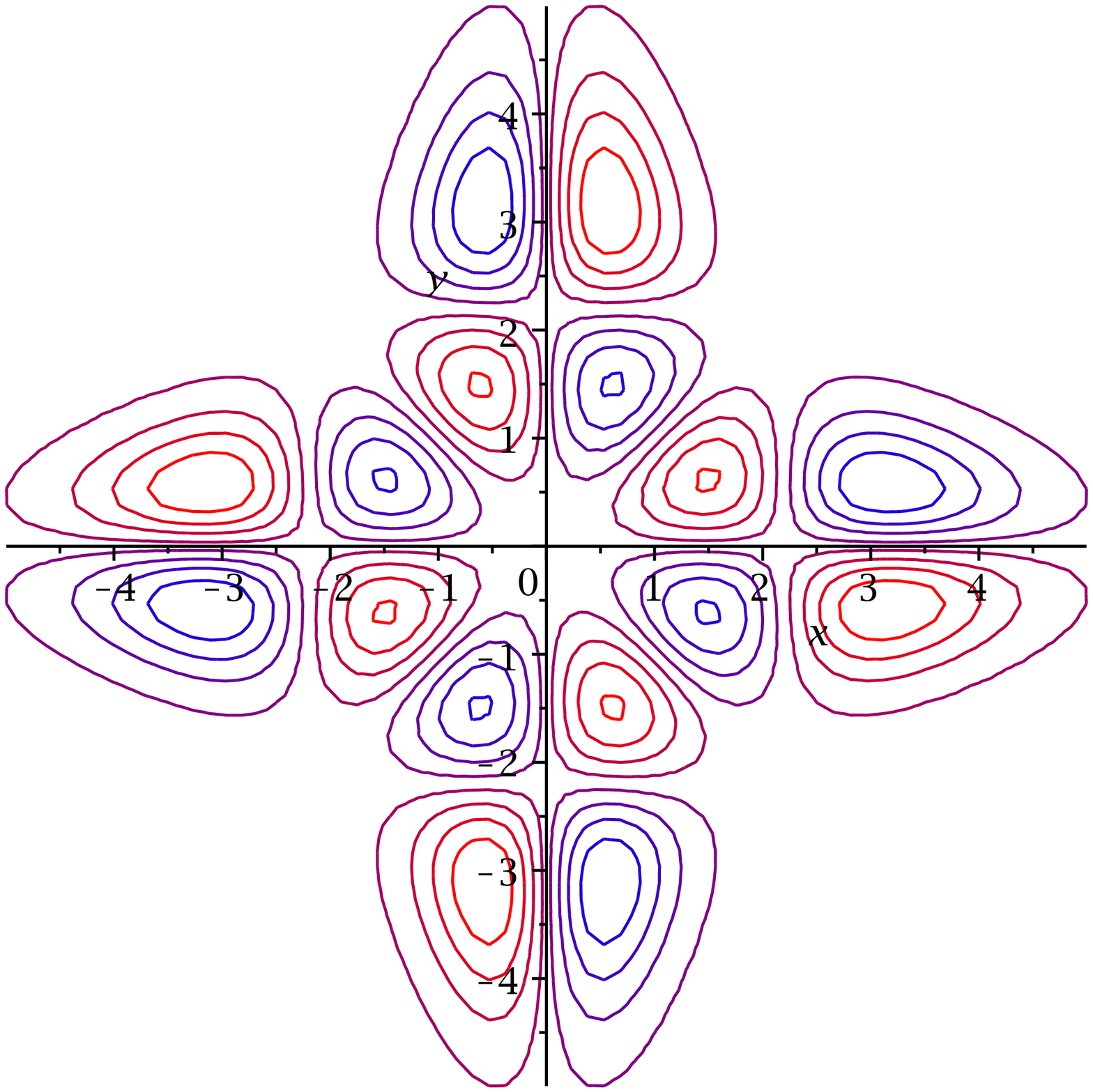}
\par
\bigskip
\end{center}
\caption{Contour lines for the eigenfunctions $1A_1$, $2A_1$, $1A_2$ and $%
2A_2$ }
\label{fig:A}
\end{figure}

\begin{figure}[H]
~\bigskip\bigskip
\par
\begin{center}
\includegraphics[width=6cm]{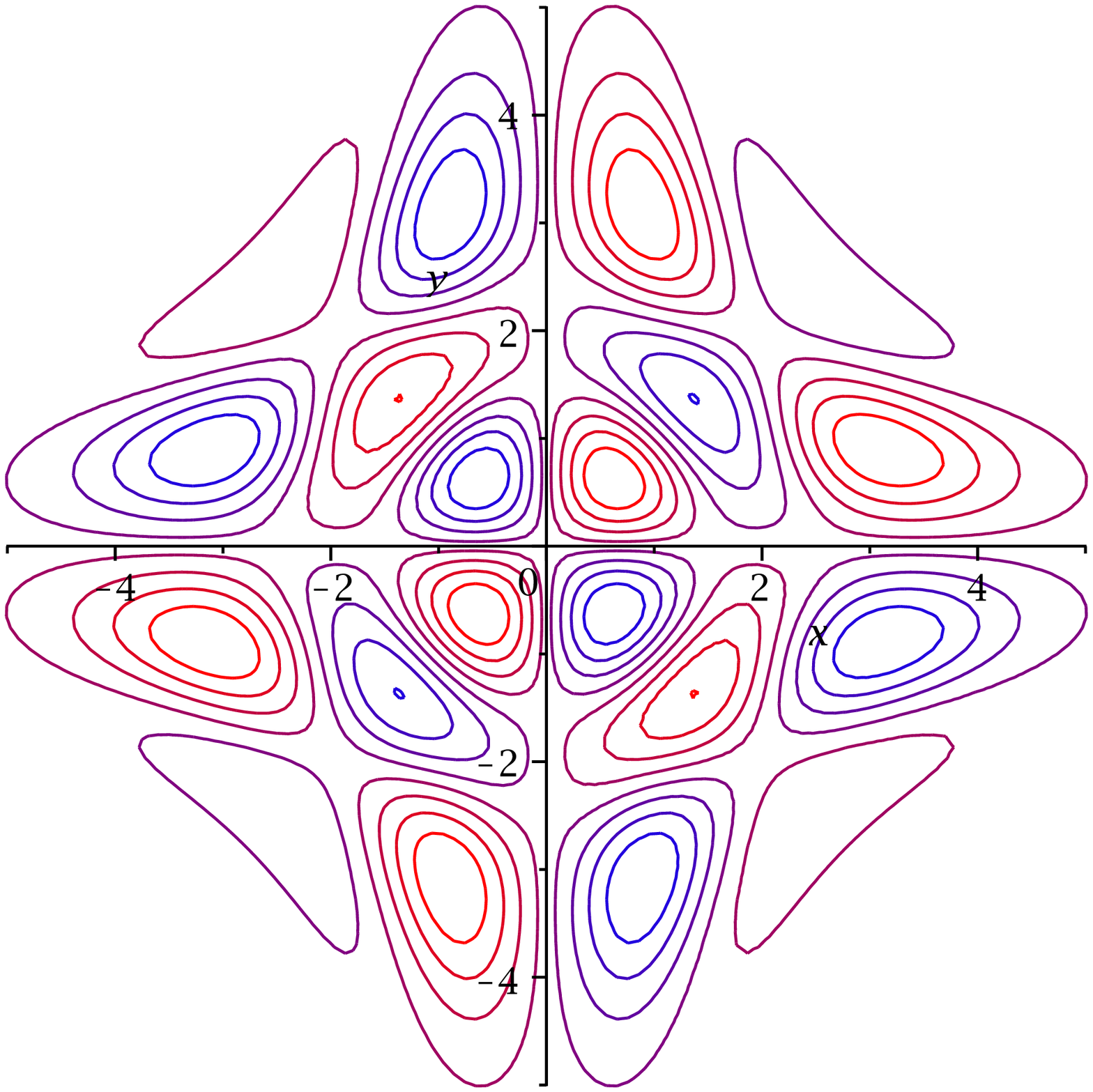} \includegraphics[width=6cm]{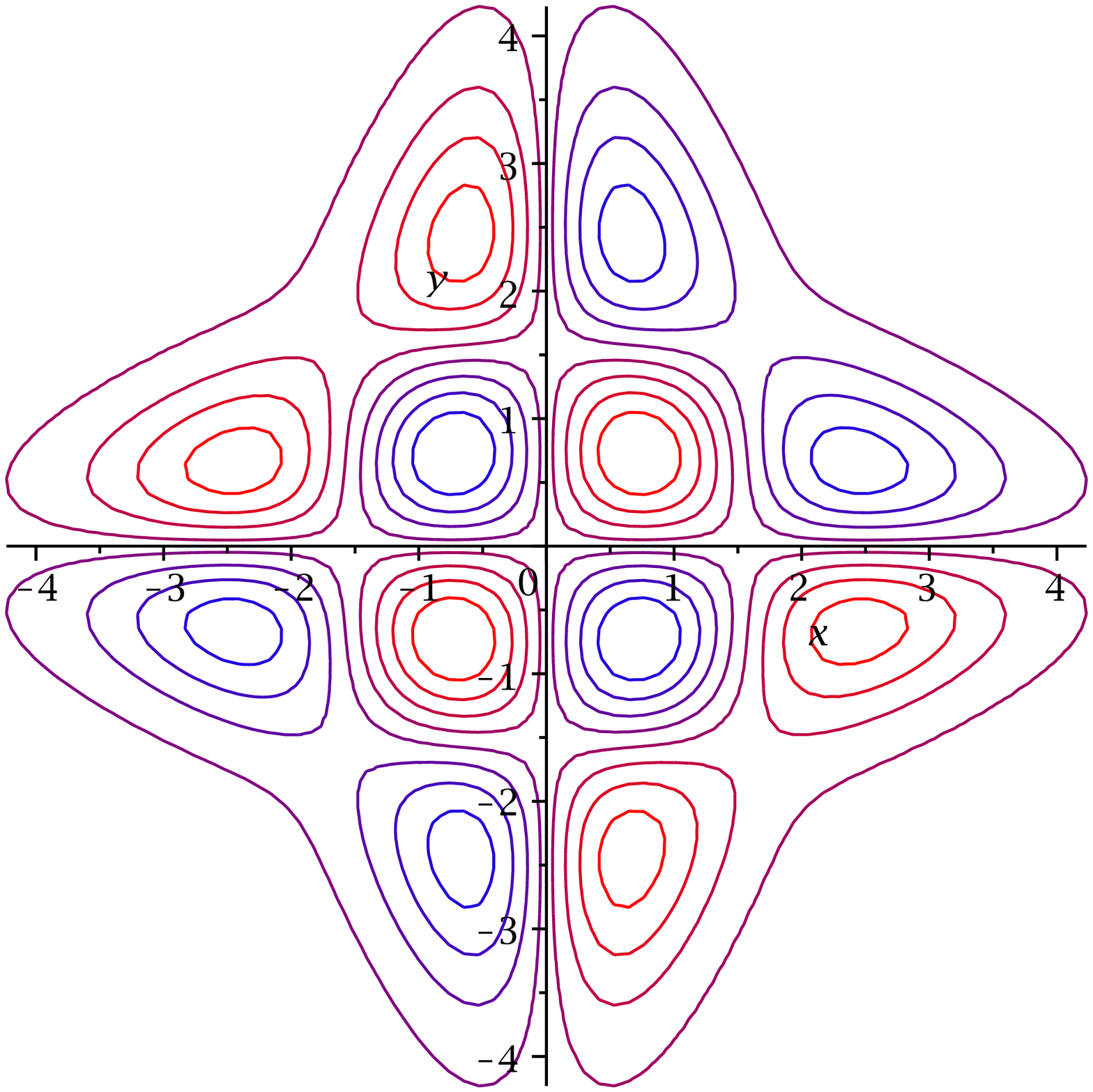} %
\includegraphics[width=6cm]{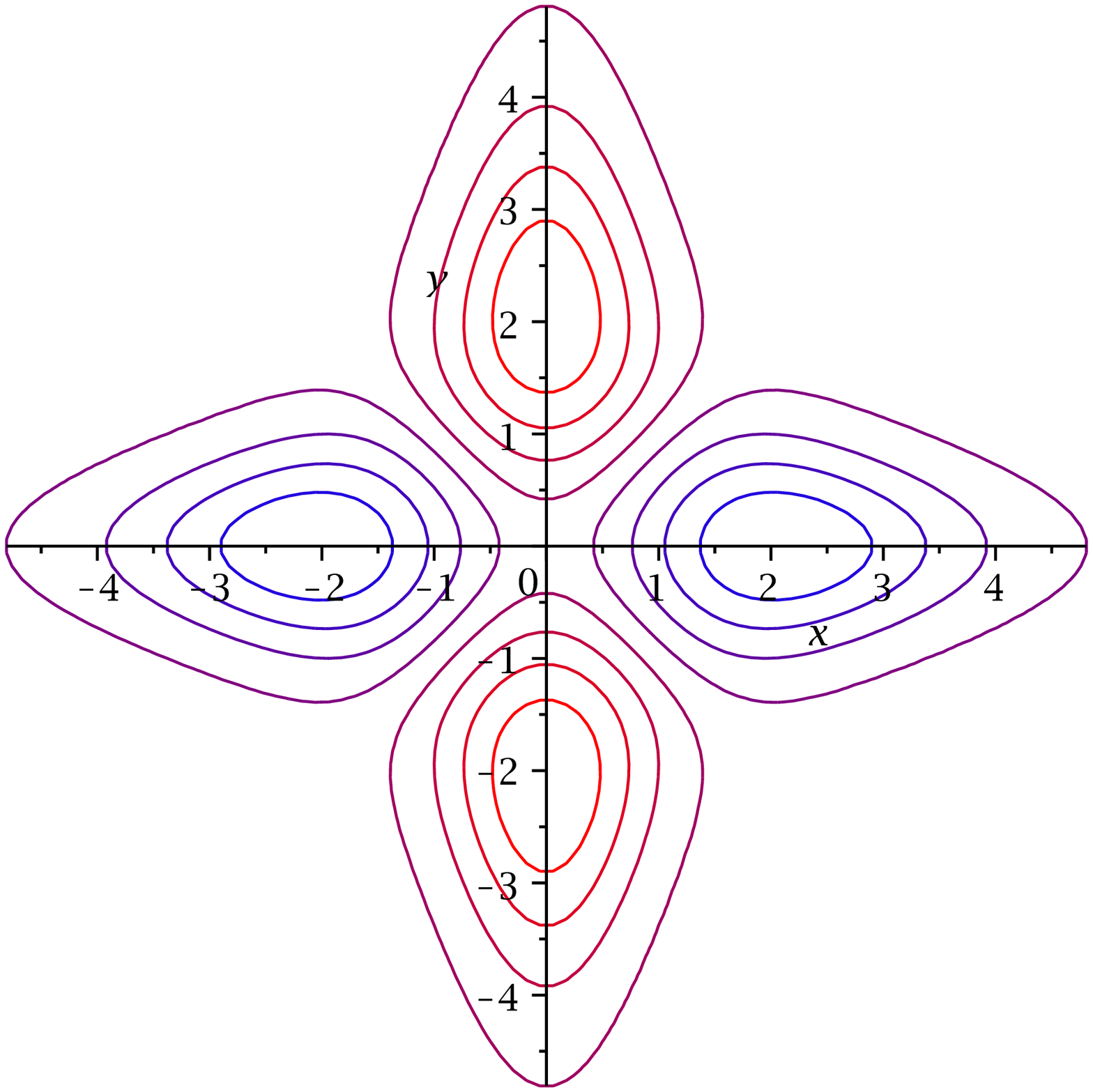}\includegraphics[width=6cm]{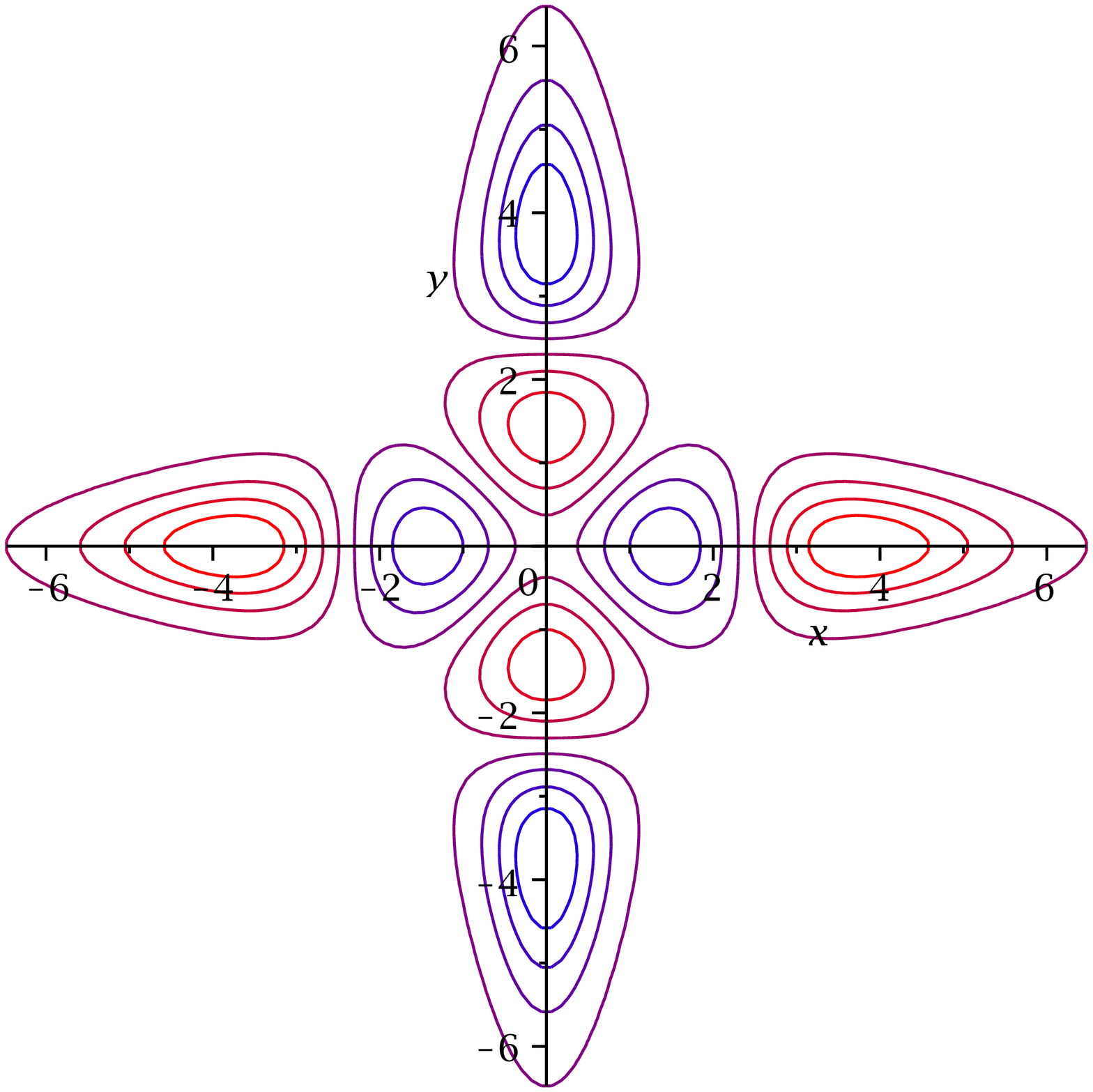}
\par
\bigskip
\end{center}
\caption{Contour lines for the eigenfunctions $1B_1$, $2B_1$, $1B_2$ and $%
2B_2$}
\label{fig:B}
\end{figure}

\begin{figure}[H]
~\bigskip\bigskip
\par
\begin{center}
\includegraphics[width=6cm]{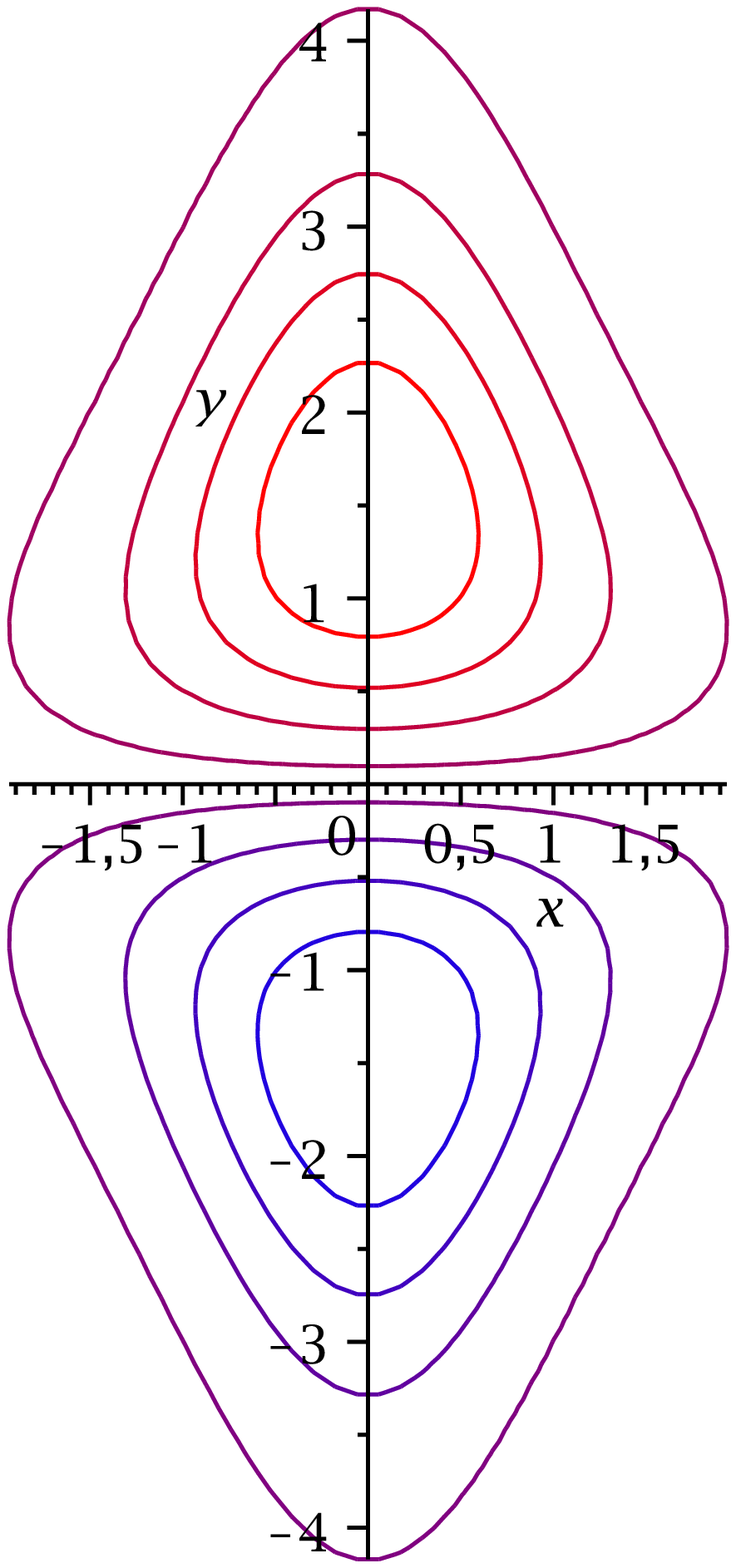} \includegraphics[width=6cm]{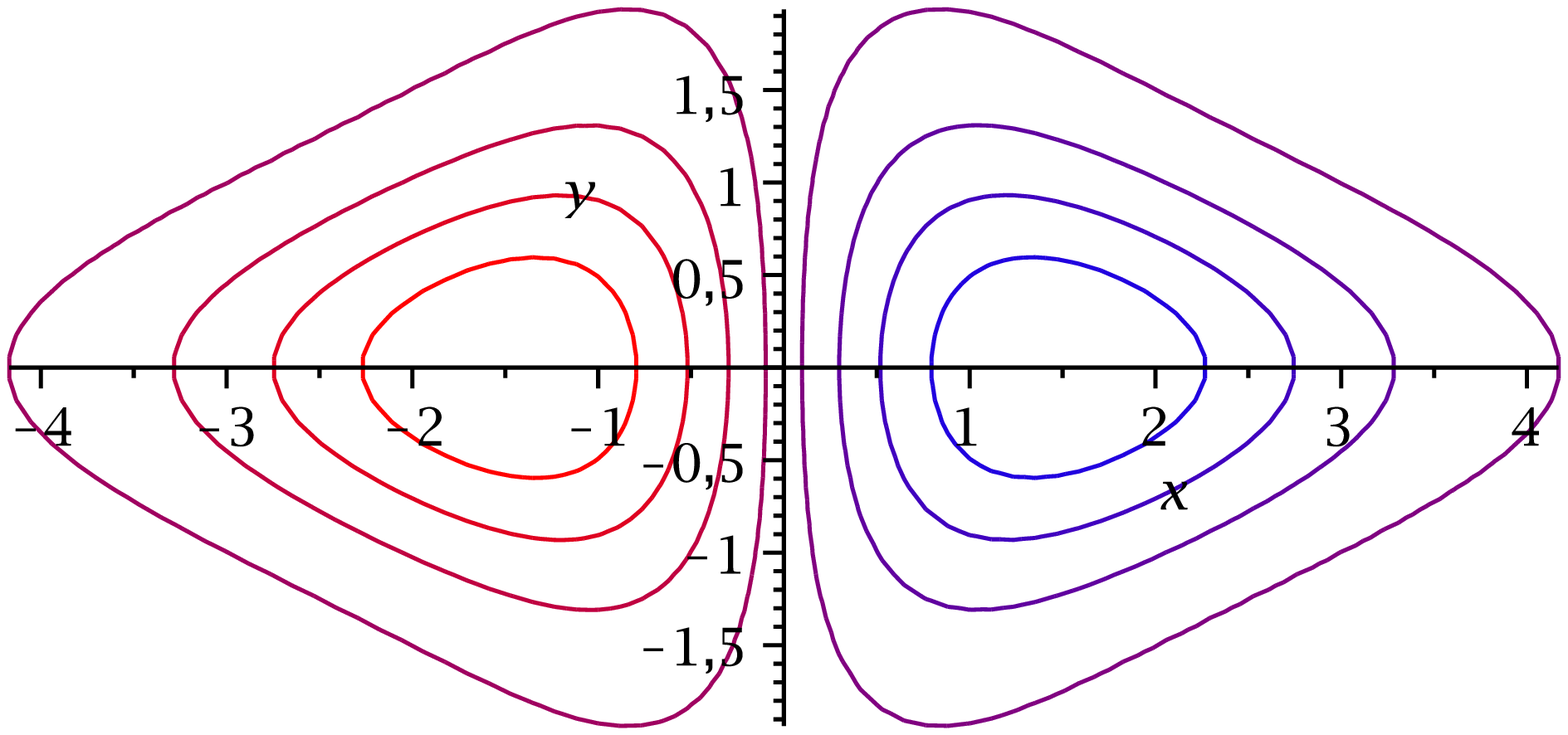}
\includegraphics[width=6cm]{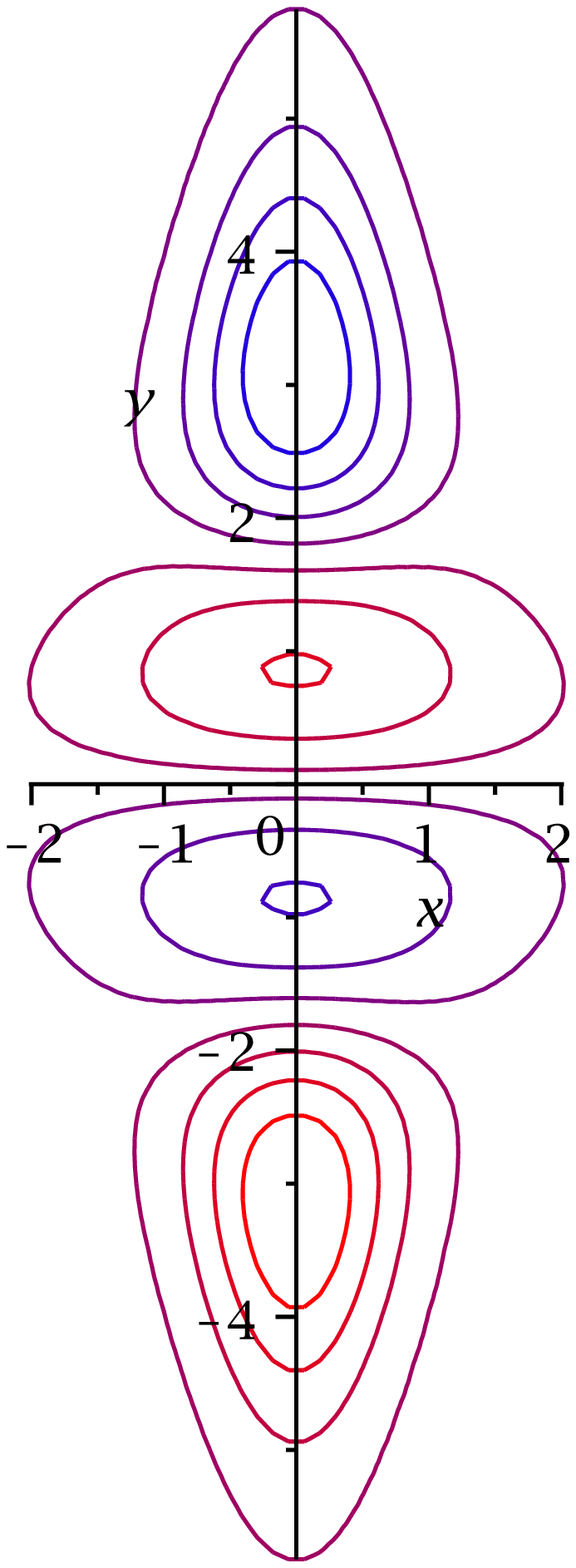}\includegraphics[width=6cm]{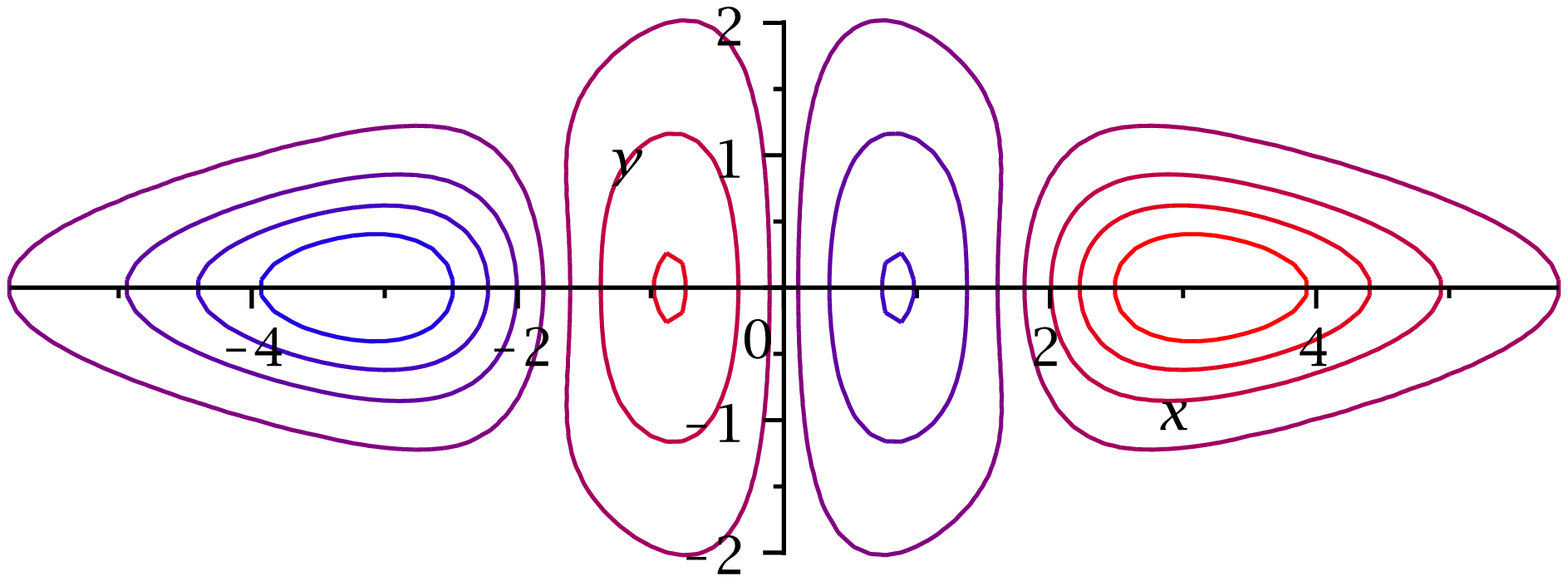}
\par
\bigskip
\end{center}
\caption{Contour lines for the two-fold degenerate eigenfunctions $1E$ and $%
2E$ }
\label{fig:E}
\end{figure}

Table~\ref{tab:E_x2y2} shows results for the lowest eigenvalues
obtained by the three methods outlined above. The RRHO ones are
the most accurate because they are based on basis sets of
dimension $D\leq 1035$. The RRK and CMX results were obtained with
smaller basis sets because their purpose is merely to verify the
RRHO results. The CMX is the less reliable of the three methods as
argued elsewhere\cite{AF09} but it is a suitable independent test
because it is not based on the variational method. It is possible
to improve the RRK and CMX results by choosing $a$ conveniently;
however, here we simply chose $a=1$ that is not optimal for all
the states. Figures \ref {fig:A}, \ref{fig:B} and \ref{fig:E} show
contour lines for some of the states of the anharmonic oscillator
obtained by means of the RRHO.

The two variational methods appear to converge rather slowly but
smoothly from above as expected for such approaches. Numerical
instabilities appeared for the greatest RRHO matrices and we
estimated the eigenvalues from the best results that satisfied the
well known variational inequality $E^{(D+m)}<E^{(D)}$. The CMX
does not give upper bounds but it approached the variational
results satisfactorily. No anomalous behaviour was detected that
could suggest that the spectrum is not discrete. Therefore,
present numerical results support the mathematical proofs given by
Simon\cite{S83} and stand against the claim raised by Bender et
al\cite{BDMS01}.

\begin{table}[H]
\caption{First eigenvalues of the anharmonic oscillator (\ref{eq:H})
calculated by means of the three methods discussed in sections \ref{sec:RRVM}
and \ref{sec:Moments}.}
\label{tab:E_x2y2}
\begin{center}
\par
\begin{tabular}{|l|D{.}{.}{12}|D{.}{.}{10}|D{.}{.}{10}|}
\hline \multicolumn{1}{|c}{State} & \multicolumn{1}{|c|}{RRHO} & \multicolumn{1}{c|}{RRK} & \multicolumn{1}{c|}{CMX}\\
\hline

$1A_1$ &  1.10822315759  &   1.108224    &    1.10822 \\
$1E  $ &  2.37863782934  &   2.37869     &    2.376   \\
$1B_1$ &  3.05608115466  &   3.0563      &    3.055   \\
$2A_1$ &  3.5149490453   &   3.518       &            \\
$2E  $ &  4.09346927636  &   4.10        &            \\
$2B_1$ &  4.75277240183  &   4.78        &            \\
$3A_1$ &  4.98496358748  &   5.07        &            \\
$1B_2$ &  5.01127928154  &   5.01127930  &    5.0112  \\
$3E  $ &  5.498979516    &   5.7         &            \\
$3B_1$ &  6.1448192750   &   6.47        &            \\
$4A_1$ &  6.237128106    &               &            \\
$4E  $ &  6.67235007     &               &            \\
$5E  $ &  7.1810983      &               &            \\
$4B_1$ &  7.37557348     &               &            \\
$5A_1$ &  7.381759978    &               &            \\
$6E  $ &  7.999          &               &            \\
$1A_2$ &  8.074373925386 &   8.0743745   &    8.0738  \\

\hline
\end{tabular}
\end{center}
\end{table}

\end{document}